\documentclass[epj,nopacs]{svjour}
\usepackage{graphics}
\begin{document}

\newcommand{\blhill}{Department of Physics, Black Hills State University, Spearfish, SD, 
USA}
\newcommand{\ITEP}{National Research Center ``Kurchatov Institute'' Institute for Theoretical and Experimental Physics, Moscow, Russia}
\newcommand{\JINR}{Joint Institute for Nuclear Research, Dubna, Russia}
\newcommand{\lbnl}{Nuclear Science Division, Lawrence Berkeley National Laboratory, Berkeley, CA, USA}
\newcommand{\lanl}{Los Alamos National Laboratory, Los Alamos, NM, USA}
\newcommand{\queens}{Department of Physics, Engineering Physics and Astronomy, Queen's University, Kingston, ON, Canada} 
\newcommand{\uw}{Center for Experimental Nuclear Physics and Astrophysics, 
and Department of Physics, University of Washington, Seattle, WA, USA}
\newcommand{\unc}{Department of Physics and Astronomy, University of North Carolina, Chapel Hill, NC, USA}
\newcommand{\duke}{Department of Physics, Duke University, Durham, NC, USA}
\newcommand{\ncsu}{Department of Physics, North Carolina State University, Raleigh, NC, USA}	
\newcommand{\ornl}{Oak Ridge National Laboratory, Oak Ridge, TN, USA}
\newcommand{\ou}{Research Center for Nuclear Physics, Osaka University, Ibaraki, Osaka, Japan}
\newcommand{\pnnl}{Pacific Northwest National Laboratory, Richland, WA, USA}
\newcommand{\ttu}{Tennessee Tech University, Cookeville, TN, USA}
\newcommand{\sdsmt}{South Dakota School of Mines and Technology, Rapid City, SD, USA}
\newcommand{\usc}{Department of Physics and Astronomy, University of South Carolina, Columbia, SC, USA}
\newcommand{\usd}{Department of Physics, University of South Dakota, Vermillion, SD, USA} 
\newcommand{\ut}{Department of Physics and Astronomy, University of Tennessee, Knoxville, TN, USA}
\newcommand{\tunl}{Triangle Universities Nuclear Laboratory, Durham, NC, USA}

\title{Search for Pauli Exclusion Principle Violating Atomic Transitions and Electron Decay with a P-type Point Contact Germanium Detector}

\author{N.~Abgrall\inst{6}	
\and I.J.~Arnquist\inst{10}
\and F.T.~Avignone~III\inst{8, 16}
\and A.S.~Barabash\inst{3}
\and F.E.~Bertrand\inst{8}
\and A.W.~Bradley\inst{6}	
\and V.~Brudanin\inst{4}
\and M.~Busch\inst{2, 13}	
\and M.~Buuck\inst{19}  
\and A.S.~Caldwell\inst{12}	
\and Y-D.~Chan\inst{6}
\and C.D.~Christofferson\inst{12} 
\and P.-H.~Chu\inst{5} 
\and C. Cuesta\inst{19}	
\and J.A.~Detwiler\inst{19}	
\and C. Dunagan\inst{12}	
\and Yu.~Efremenko\inst{18}
\and H.~Ejiri\inst{9}
\and S.R.~Elliott\inst{5}
\and P.S~Finnerty\inst{13, 15}
\and A.~Galindo-Uribarri\inst{8}	
\and T.~Gilliss\inst{13, 15}  
\and G.K.~Giovanetti\inst{13, 15} \thanks{email: gkg@princeton.edu}	  
\and J. Goett\inst{5}	
\and M.P.~Green\inst{7, 8, 13}   
\and J. Gruszko\inst{19}		
\and I.S.~Guinn\inst{19}		
\and V.E.~Guiseppe\inst{16}	
\and R.~Henning\inst{13, 15}
\and E.W.~Hoppe\inst{10}
\and S. Howard\inst{12}  
\and M.A.~Howe\inst{13, 15}
\and B.R.~Jasinski\inst{17}  
\and K.J.~Keeter\inst{1}
\and M.F.~Kidd\inst{14}	
\and S.I.~Konovalov\inst{3}
\and R.T.~Kouzes\inst{10}
\and B.D.~LaFerriere\inst{10}   
\and J. Leon\inst{19}	
\and J.~MacMullin\inst{13, 15} 
\and R.D.~Martin\inst{11}	
\and R. Massarczyk\inst{5}		
\and S.J.~Meijer\inst{13, 15}	
\and S.~Mertens\inst{6}		
\and J.L.~Orrell\inst{10} 
\and C. O'Shaughnessy\inst{13, 15}	
\and A.W.P.~Poon\inst{6}
\and D.C.~Radford\inst{8}
\and J.~Rager\inst{13, 15}	
\and K.~Rielage\inst{5}
\and R.G.H.~Robertson\inst{19}
\and E. Romero-Romero\inst{8, 18} 
\and B.~Shanks\inst{13, 15}	
\and M.~Shirchenko\inst{4}
\and A.M.~Suriano\inst{12} 
\and D.~Tedeschi\inst{16}		
\and J.E.~Trimble\inst{13, 15}	
\and R.L.~Varner\inst{8}  
\and S. Vasilyev\inst{4}	
\and K.~Vetter\inst{6} \thanks{Alternate address: Department of Nuclear Engineering, University of California, Berkeley, CA, USA}
\and K.~Vorren\inst{13, 15} 
\and B.R.~White\inst{8}	
\and J.F.~Wilkerson\inst{8, 13, 15}    
\and C. Wiseman\inst{16}		
\and W.~Xu\inst{13, 15} 
\and E.~Yakushev\inst{4}
\and C.-H.~Yu\inst{8}
\and V.~Yumatov\inst{3}
\and I.~Zhitnikov\inst{4}
} 

\institute{
\blhill 
\and \duke
\and \ITEP
\and \JINR
\and \lanl
\and \lbnl
\and \ncsu
\and \ornl
\and \ou
\and \pnnl
\and \queens
\and \sdsmt
\and \tunl
\and \ttu
\and \unc
\and \usc
\and \usd
\and \ut
\and \uw}

\date{Received: date / Revised version: date}
%
\abstract{
A search for Pauli-exclusion-principle-violating K$_{\alpha}$ electron transitions was performed using \mbox{89.5~kg-d} of data collected with a p-type point contact high-purity germanium detector operated at the Kimballton Underground Research Facility. A lower limit on the transition lifetime of $5.8\times10^{30}$~seconds at 90\%~C.L. was set by looking for a peak at 10.6~keV resulting from the x-ray and Auger electrons present following the transition. A similar analysis was done to look for the decay of atomic K-shell electrons into neutrinos, resulting in a lower limit of $6.8\times10^{30}$~seconds at 90\%~C.L.  It is estimated that the \textsc{Majorana Demonstrator}, a 44~kg array of p-type point contact detectors that will search for the neutrinoless double-beta decay of $^{76}$Ge, could improve upon these exclusion limits by an order of magnitude after three years of operation.
} 
%
\maketitle
\section{Introduction}
\label{sec:intro}
In 1925, Wolfgang Pauli postulated the exclusion principle to describe the periodic nature of the elements~\cite{pau25}. The Pauli exclusion principle (PEP) states that there can never be two or more equivalent electrons in an atom, i.e. no two identical fermions can occupy the same quantum state. In quantum field theory, PEP emerges as a consequence of the application of the spin-statistics theorem to anti-commuting fields and is considered a fundamental law of nature. However, the discovery of parity non-conservation in 1957 spurred a new set of experiments testing fundamental laws~\cite{wu57}.  This included a search for PEP-violating transitions of atomic electrons in iodine by Reines and Sobel, who in 1974 gave a lower limit on the lifetime of non-Paulian transitions of $2 \times 10^{27}$~seconds~\cite{rei74}.  Following this pioneering work, many searches have been done for atoms and nuclei in PEP-violating states~\cite{ell12}.  

Despite early efforts to construct models that incorporate PEP violation~\cite{ign87,gre87}, there is no established theoretical framework for Pauli exclusion principle violation. The most promising theory that allows small violations of Fermi statistics is quon theory~\cite{gre91}, which, despite many favorable properties, violates locality~\cite{gre00}. The situation is further complicated by the Messiah-Greenberg superselection rule, which states that if there is a small mixed symmetry component in a primarily antisymmetric wave-function, the Hamiltonian would only connect a mixed state to another mixed state. This means that even if the exclusion principle is violated, PEP violating transitions of electrons or nucleons to lower orbitals would still be forbidden~\cite{mes64,ama80}.

In light of this constraint, Elliott et al. developed a scheme for categorizing PEP violation experiments based on the ``newness'' of the fermion in the system~\cite{ell12}, i.e.~the time at which the fermion initially interacts with the nucleus or atom and establishes the symmetry of the wave-function. Type-I experiments involve a fermion that has never interacted with another fermionic system. This could be a fermion created during the big bang that forms an anomalous nuclear state~\cite{tho92} or an electron emitted during $\beta$ decay or pair production that undergoes a PEP violating capture by an atomic system~\cite{gol48}. Type-II experiments use fermions that have previously formed wave-functions with other fermions but have never interacted with the system under examination. Ramberg and Snow~\cite{ram90} and Elliott et al.~\cite{ell12} perform Type-II experiments by running a current through a conductor and looking for PEP violating electron captures of electrons from the current source. Type-III experiments look for the PEP violating transition of an atomic electron or nucleon in an existing system, where the symmetry of the wave-function has already been established. A theoretical description of this type of PEP violation contradicts the Messiah-Greenberg superselection rule or requires the use of extra-dimensions, electronic substructure, or other exotic physics~\cite{aka92,gre89}. 

The results from PEP violation experiments can be compared using the parameter $\frac{1}{2}\beta^2$, which measures the probability that two fermions form a state with a symmetric wave function component. As Ref.~\cite{ell12} points out, this is a simplistic method of comparing experiments and warrants the wide variety of approaches used to search for PEP violation. A comprehensive table of existing experimental limits and their classification is given in Ref.~\cite{ell12}.

The most stringent constraint on Type-III PEP violations in atomic transitions comes from the DAMA/LIBRA experiment, a 250~kg array of radio-pure NaI(Tl) detectors at the Gran Sasso National Laboratory. DAMA/LIBRA searched for PEP violating K-shell transitions in iodine using 0.53~ton-y of data and set a lower limit on the transition lifetime of $4.7\times10^{30}$~seconds and constrained $\frac{1}{2}\beta^2 < 1.28\times10^{-47}$ at 90\% C.L.~\cite{ber09}. A similar Type-III experiment can be performed using a high-purity germanium (HPGe) detector with an energy threshold sufficiently low to observe the transition of an L-shell germanium electron into an already full K-shell. This event would deposit roughly the same energy as a standard K$_{\alpha}$ transition.  However, due to the increased shielding of the nuclear charge from the second electron in the K-shell, the energy of the transition would be shifted down in energy by a few hundred electronvolts. M. Chen has calculated the x-ray energy of a PEP violating K-shell transition in germanium to be 9.5 keV~\cite{ell12}. When measuring this decay with a HPGe detector, the x-ray energy sums with the emissions from the further relaxation of the atomic shells creating a feature at 10.6 keV, on the high energy shoulder of the $^{68}$Ge K-shell capture line. Because the probabilities of the x-ray and the associated relaxation emissions escaping the detector are vanishingly small, the efficiency for detecting a PEP violating K$_\alpha$ transition occurring within the detector active volume is effectively 100\%.

\section{MALBEK}
\label{sec:malbek}
The \textsc{Majorana} collaboration is currently constructing the \textsc{Majorana Demonstrator}, a 44~kg array of p-type point contact (PPC) HPGe detectors that will search for the neutrinoless double-beta decay of $^{76}$Ge. Due to their small capacitance, PPC detectors can be operated with sub-keV energy thresholds, making them capable of searching for a PEP violating K-shell transition. The \textsc{Majorana} Low-background Broad Energy Germanium Detector at KURF (MALBEK) is a 450~g PPC that operated at the Kimballton Underground Research Facility (KURF) in Ripplemeade, VA. MALBEK was used to explore sources of background and the performance of PPC detectors in the K-shell transition region of interest in order to establish the sensitivity of the \textsc{Demonstrator} to physics in that region. A complete description of the MALBEK detector can be found in Refs.~\cite{fin13,gio14,gio15}. 

The MALBEK detector began collecting shielded data at KURF on 15 November 2011. Data taking proceeded for 288 days, ending on 8 August 2012. Due to a period of frequent power outages at KURF, the dataset is divided into two distinct run periods, 15 November 2011 to 12 March 2012 and 9 April 2012 to 29 August 2012. There were additional intervals of down-time within the two run periods caused by intermittent power outages at the mine, reducing the total live-time of the detector to 221.5 days.

\section{Analysis}
\label{sec:analysis}
Analysis of digitized waveform data from MALBEK is done using the Germanium Analysis Toolkit (\texttt{GAT}), a modular data analysis framework developed by the \textsc{Majorana} collaboration. After the data are processed, a set of basic data selection cuts are applied. First, periods of high noise are removed from the dataset, e.g. data collected immediately following a power outage. Then a set of timing cuts are performed to remove events coincident with preamplifier reset inhibit pulses and events occurring within 15 minutes of a liquid nitrogen dewar fill. Finally, cuts based on the waveform shape are applied to eliminate non-physics signals caused by microphonics and bias voltage micro-discharges.

In addition to the basic data selection cuts, a cut is applied to remove events that originate near the detector surface. The n+ surface contact on the MALBEK detector is created by diffusing lithium into the crystal lattice, resulting in an approximately 0.5 - 1 mm thick region of n+ material extending into the bulk of the crystal. Because of the high impurity concentration in this region, much of the n+ contact volume remains un-depleted when the detector is biased. However, some fraction of the charge created by an interaction occurring within the contact can diffuse into the depleted region of the crystal and induce a signal in the same manner as a bulk interaction. The amplitude of this signal will only reflect the fraction of initial charge carriers that moved into the depleted region, and the full energy of the originating interaction will be lost. In this way, surface events from higher energy sources become a background in the PEP-violating signal region of interest.

\begin{figure}
\resizebox{0.5\textwidth}{!}{%
  \includegraphics{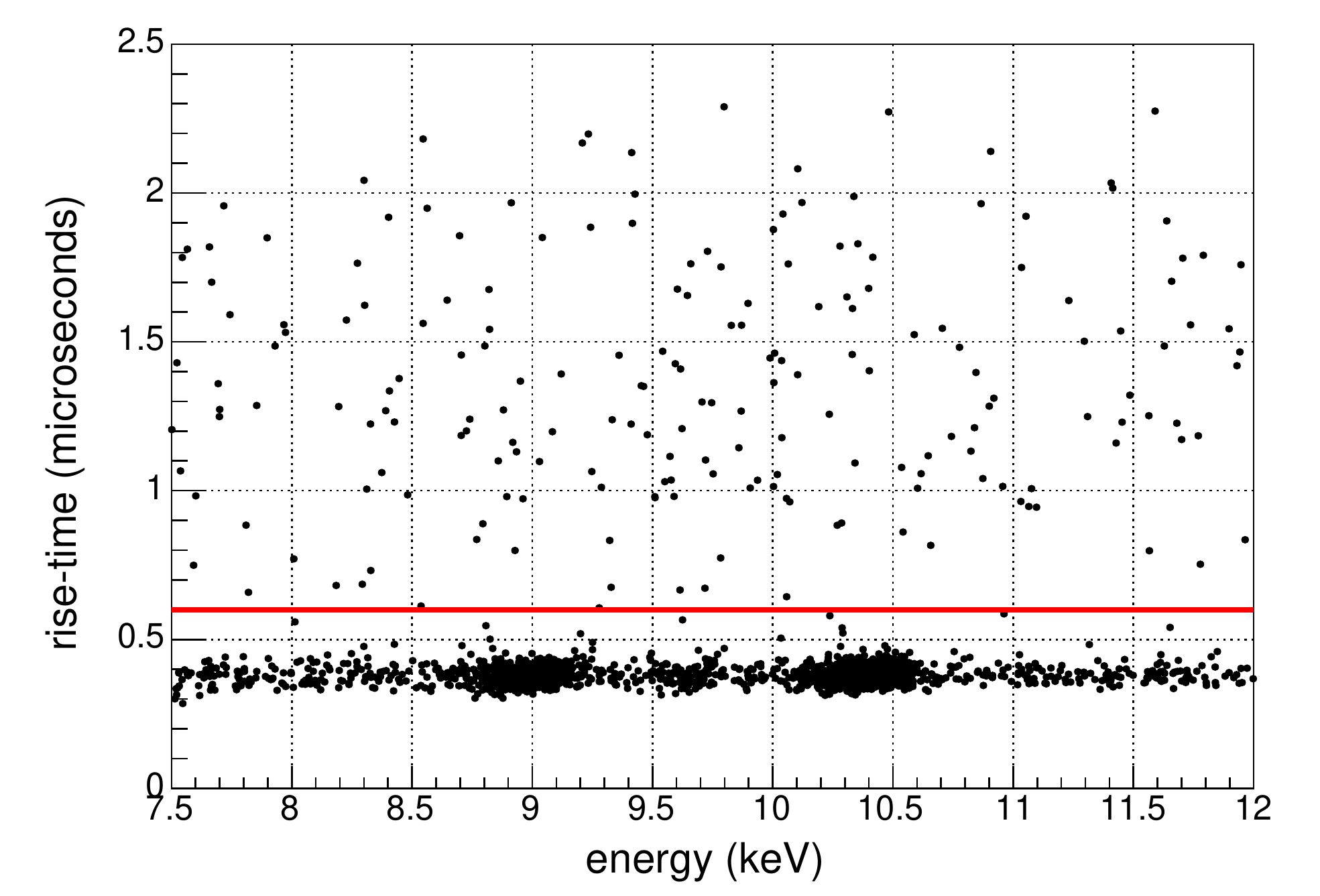}
}
\caption{Rise-time distribution of events in the region of interest. The rise-time cut, defined to retain more than 99\% of all fast bulk events, is shown by the red line.}
\label{fig:pep_risetime_dist}       
\end{figure}

Energy degraded surface events can be distinguished from bulk events by measuring their rise-time, the time it takes the charge signal to reach its maximum value. Holes created in the n+ contact take microseconds to diffuse into the bulk of the detector and induce a signal, resulting in charge signals with correspondingly long rise-times~\cite{agu13}. This is in sharp contrast to events that occur in the detector bulk, where all holes are collected within several hundred nanoseconds. A cut based on the charge signal rise-time was implemented to remove surface events from the MALBEK dataset. The cut was calibrated using pulser-generated data to remove slow surface events with high efficiency while retaining more than 99\% of all fast bulk events. As shown in Figure~\ref{fig:pep_risetime_dist}, the slow surface event and fast bulk event populations in the energy region around the PEP-violating decay peak are clearly separated in rise-time, and the uncertainty on the signal acceptance due to this cut is negligible. A small number of surface events may pass the cut, but with no a priori expectation for the surface event distribution, no correction is made for these events.

The search for PEP-violating K-shell transitions within the MALBEK detector is performed using an unbinned maximum likelihood analysis to data that fall between 7.5 and 12.0~keV. The dominant backgrounds at 10.6~keV are the high-energy tail of the 10.37~keV K-shell capture line from cosmogenically activated $^{68}$Ge, and a featureless continuum due primarily to the forward Compton scattering of gamma-rays emitted by primordial contaminants ($^{238}$U, $^{232}$Th, and $^{40}$K) as well as cosmogenic cobalt isotopes in MALBEK detector components. The 7.5 to 12.0~keV region also includes K-capture peaks from $^{65}$Zn (8.98~keV) and $^{68}$Ga (9.66~keV). If the energy region containing the $^{65}$Zn and $^{68}$Ga peaks is not used in the fit, the rate in the $^{68}$Ge peak is not well constrained and the quality of the fit suffers, which in turn reduces the sensitivity of the experiment. Increasing the fit region beyond 7.5 or 12~keV does not improve the sensitivity further. 

The background model used in the analysis includes a flat continuum and the $^{65}$Zn, $^{68}$Ga, and $^{68}$Ge K-shell capture peaks. Each peak is described by a Gaussian and a smoothed step function that increases the background continuum on the low-energy side of the peak, e.g. to account for inefficiency of the surface event cut. The relative amplitude of the smoothed step functions to the peak amplitudes is included as a single fit parameter. The signal model consists of a single Gaussian at 10.6~keV. The widths of the signal peak and the background K-shell capture peaks are constrained to follow the function $\sigma(E) = (a+bE)^{\frac{1}{2}}$, where $\sigma(E)$ is the peak width at energy $E$ and $a$ and $b$ are allowed to float during the fit.

The detector energy scale was determined to be linear at better than a percent using the $^{68}$Ge L-shell capture line (1.30~keV), the $^{55}$Fe K-shell capture line (6.54~keV), and the $^{65}$Zn K-shell capture line. During the fit, the relative peak positions of the three background peaks and the signal peak are constrained to be linear, but the calibration parameters are allowed to float. 

The dominant uncertainty in the detector exposure arises from the uncertainty in the detector's active volume. This was calculated by comparing the ratio of events observed in the 81 keV and 356 keV peaks from a $^{133}$Ba source run to a detailed Monte Carlo simulation~\cite{agu13}. It was determined that the full charge collection depth within the detector is $933\pm120$ $\mu$m. This reduces the active mass of the detector from 465 g to $404.2 \pm 15$ g and results in a total exposure of $89.5\pm3.3$ kg-d. The normalization of the signal peak is allowed to float within a 3.7\% Gaussian constraint to incorporate this source of uncertainty.

\section{Results}
\label{sec:results}
The best fit value for the PEP violating peak, which is allowed to float to unphysical values to avoid discontinuities in the profile likelihood ratio at the parameter boundary, is -0.22~counts/kg-d. Following the method described in Ref.~\cite{rol05}, the number of PEP events is set to zero and the maximum likelihood value at this point is used to construct the profile likelihood ratio. With no systematic uncertainties included in the likelihood function, a signal event rate greater than 0.11~counts/kg-d is excluded at the 90\% C.L. Including uncertainties on the detector efficiency in the likelihood function results in a slightly weaker 90\% C.L. exclusion of 0.12~counts/kg-d. A Monte Carlo calculation over an
ensemble of identical experiments assuming the best-fit energy spectrum without a signal yields a sensitivity of 0.27~counts/kg-d. The probability of obtaining a limit stronger than the presented result is 10\%.

\begin{figure}
\resizebox{0.5\textwidth}{!}{%
  \includegraphics{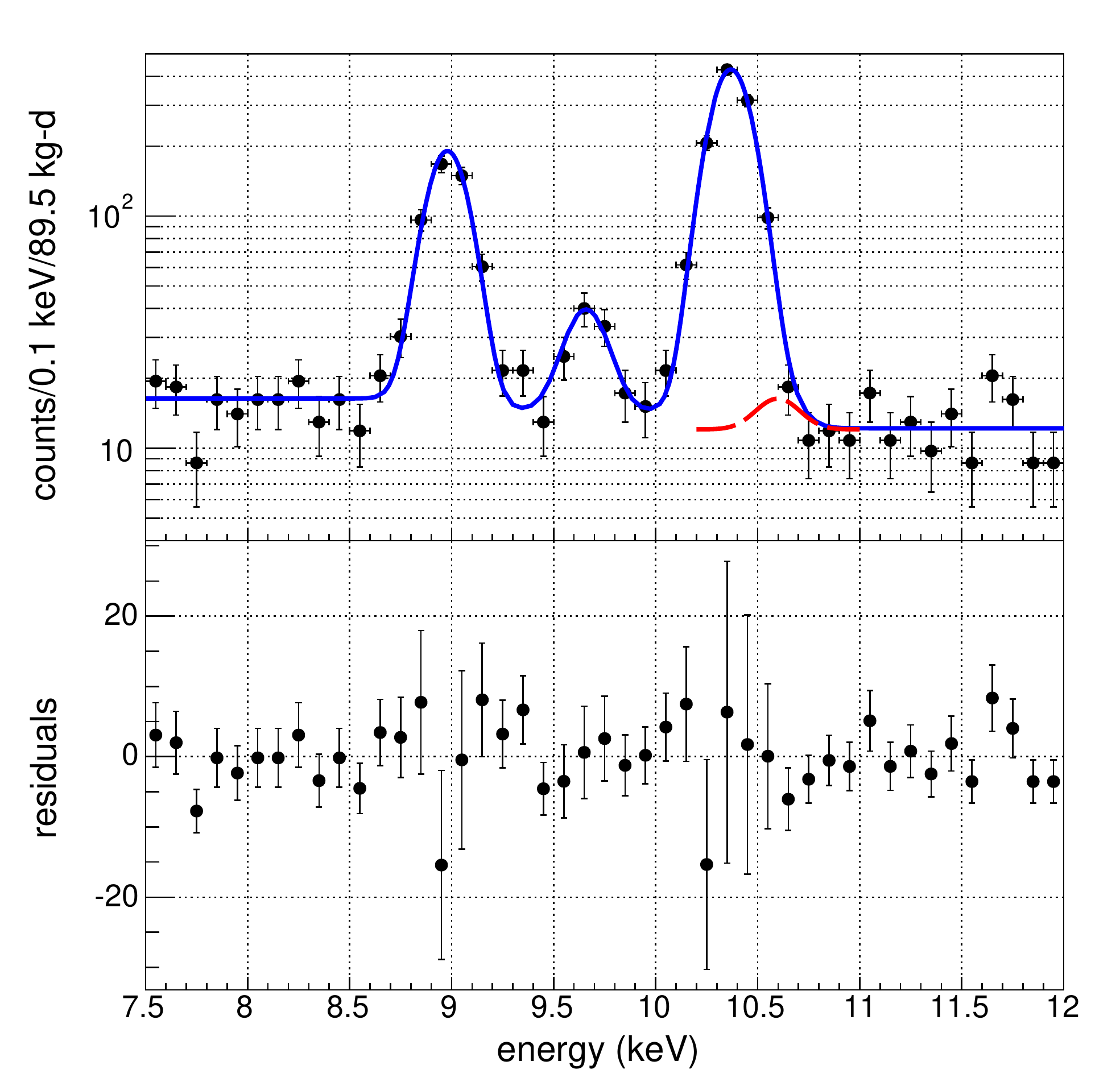}
}
\caption{Fit of the signal and background model (blue) to \mbox{89.5~kg-d} of MALBEK data (black points). The number of events in the PEP-violating decay peak is fixed at the the 90\% C.L. exclusion limit, 0.12 counts/kg-d. The PEP-violating signal is shown separately on top of the background flat continuum (red dashed). Fit residuals are shown in the bottom panel.}
\label{fig:pep_90_fit}       
\end{figure}

The best fit of the constrained model to the data with the number of events in the PEP-violating decay peak fixed at the 90\% C.L. is shown in Figure~\ref{fig:pep_90_fit}. This result corresponds to a PEP violating K$_{\alpha}$ transition lifetime of $5.8\times10^{30}$~seconds at a 90\% C.L., comparable to the limit reported by the DAMA/LIBRA collaboration of $4.7\times10^{30}$~seconds for iodine K-shell transitions, despite a much lower exposure time. The MALBEK detector is competitive with this result due to its significantly better energy resolution at the region of interest, 9.4\% for DAMA/LIBRA versus 1.0\% for MALBEK. To enable a comparison to other types of experiments searching for PEP violating states, $\frac{1}{2}\beta^2$ can be calculated by comparing the PEP transition lifetime to the  $1.7\times10^{-16}$~second lifetime of a standard K$_{\alpha}$ transition in germanium, resulting in a limit on $\frac{1}{2}\beta^2 < 2.92\times10^{-47}$. This is not as strong as the limit from DAMA/LIBRA ($\frac{1}{2}\beta^2 < 1.28\times10^{-47}$) due to the smaller MALBEK exposure and the faster allowed K$_{\alpha}$ transition rate of the higher Z iodine nuclei used by DAMA/LIBRA. Type-III experiments looking for PEP violating nuclear transitions are many orders of magnitude more sensitive, see for example~\cite{ber09}.

A similar analysis can be performed using MALBEK to test the conservation of electric charge by searching for the decay of atomic electrons within the germanium crystal into neutrinos ($e^-\rightarrow\nu_e\bar{\nu}_e\nu_e$)~\cite{fei59}. Such an event would result in an x-ray and a subsequent atomic cascade that deposits the binding energy of the decaying electron in the detector. Because of its sub-keV energy threshold, MALBEK is sensitive to the decay of K-shell electrons (11.1~keV) and L-shell electrons (1.414, 1.248, and 1.217~keV). However, the L-shell decay peaks overlap in energy with L-capture peaks from cosmogenically produced $^{65}$Zn (1.10~keV) and $^{68}$Ge (1.30~keV) that are prominent in the MALBEK spectrum~\cite{gio14}. An analysis that only considers K-shell electrons excludes a count rate larger than 0.21 counts/kg-d at the 90\% C.L. This results in a lower limit on the electron decay lifetime of $6.8\times10^{30}$~seconds at the 90\% C.L. This result is approximately 11 times worse than the best lower limit on atomic electron disappearance, also from the DAMA/LIBRA group~\cite{bel99}. Including the L-shell electron decay peaks in the analysis does not increase our experimental sensitivity due to the background from the cosmogenic L-capture peaks, despite the increased statistics from the eight additional atomic electrons.

When completed, the \textsc{Majorana Demonstrator} will operate 44 kgs of PPC detectors, 30 kgs of which are constructed from material enriched to greater than 87\% in $^{76}$Ge. Because cosmogenically produced $^{68}$Ge and $^{60}$Co are a background to the $^{76}$Ge neutrinoless double-beta decay signal of interest, the cosmic-ray exposure of the enriched material used to build the PPC detectors was carefully limited during detector fabrication. Consequently, the \textsc{Demonstrator} will have a significantly lower rate in the $^{68}$Ge K-shell capture peak compared to the MALBEK detector. Conservatively assuming a factor of 20 reduction in background rate within the PEP violating K$_{\alpha}$ transition region of interest relative to the MALBEK detector, the \textsc{Demonstrator} would have comparable sensitivity to DAMA/LIBRA with just 3~kg-y of exposure. An order of magnitude improvement in the half-life sensitivity could be expected after operating the array for three years, collecting 90~kg-y of enriched detector data. A similar order of magnitude improvement would be expected for the decay of atomic electrons into neutrinos and an additional factor of eight improvement is possible if the \textsc{Demonstrator} energy thresholds are low enough to observe the L-shell decay peaks.

\begin{acknowledgement}
This material is based upon work supported by the U.S. Department of Energy, Office of Science, Office of Nuclear Physics under Award  Numbers DE-AC02-05CH11231, DE-AC52-06NA25396, DE-FG02-97ER41041, DE-FG02-97ER41033, DE-FG02-97ER41042, DE-SC0012612, DE-FG02-10ER41715, DE-SC0010254, and DE-FG02-97ER41020. We acknowledge support from the Particle Astrophysics Program and Nuclear Physics Program of the National Science Foundation through grant numbers PHY-0919270, PHY-1003940, 0855314, PHY-1202950, MRI 0923142 and 1003399; the Russian Foundation for Basic Research, grant No. 15-02-02919; and the U.S. Department of Energy through the LANL/LDRD Program. We thank our hosts and colleagues at the Kimballton Underground Research Facility, Virginia Polytechnic Institute, and the Triangle University Nuclear Laboratories for their support and assistance with remote detector operations.

\end{acknowledgement}
%


\end{document}